\documentclass[prx,twocolumn,english,superscriptaddress,floatfix,longbibliography]{revtex4-2}
\usepackage{dcolumn}
\usepackage[utf8]{inputenc}
\usepackage{amsmath}
\usepackage{amsfonts}
\usepackage{amssymb}
\usepackage{graphicx}
\usepackage{hyperref}
\usepackage{xcolor}
\usepackage{array}
\usepackage{booktabs}
\hypersetup{colorlinks=true,linkcolor=blue,anchorcolor=blue,
            citecolor=blue,filecolor=blue,urlcolor=blue}

\newcommand{\ket}[1]{\ensuremath{\left|{#1}\right\rangle}}

\definecolor{darkrule}{gray}{0.02}

\begin{document}


\title{High-Dimensional Quantum Key Distribution via full Core-mode Encoding over Deployed Multicore Fibers}

\author{G.~H.~dos~Santos}
\email{These authors equally contributed to this work}
\affiliation{Departamento de F\'{\i}sica, Universidad de Concepci\'on,
             160-C Concepci\'on, Chile}
\affiliation{Millennium Institute for Research in Optics,
             Universidad de Concepci\'on, 160-C Concepci\'on, Chile}

\author{K.~B.~Sawada}
\email{These authors equally contributed to this work}
\affiliation{Departamento de F\'{\i}sica, Universidad de Concepci\'on,
             160-C Concepci\'on, Chile}
\affiliation{Millennium Institute for Research in Optics,
             Universidad de Concepci\'on, 160-C Concepci\'on, Chile}

\author{N.~Villalba}
\affiliation{Departamento de F\'{\i}sica, Universidad de Concepci\'on,
             160-C Concepci\'on, Chile}
\affiliation{Millennium Institute for Research in Optics,
             Universidad de Concepci\'on, 160-C Concepci\'on, Chile}

\author{C.~Jara}
\affiliation{Departamento de F\'{\i}sica, Universidad de Concepci\'on,
             160-C Concepci\'on, Chile}
\affiliation{Millennium Institute for Research in Optics,
             Universidad de Concepci\'on, 160-C Concepci\'on, Chile}

\author{N.~Guerrero}
\affiliation{Departamento de F\'{\i}sica, Universidad de Concepci\'on,
             160-C Concepci\'on, Chile}
\affiliation{Millennium Institute for Research in Optics,
             Universidad de Concepci\'on, 160-C Concepci\'on, Chile}

\author{C.~Melo}
\affiliation{Departamento de Ingenier\'{\i}a El\'ectrica,
             Universidad de Concepci\'on, 160-C Concepci\'on, Chile}

\author{M. H. Magiotto}
\affiliation{Instituto de Física, Universidade Federal de Goiás, Goiânia, Goiás 74.690-900, Brazil}

\author{D.~Martinez}
\affiliation{University of Vienna, Faculty of Physics,
             Vienna Center for Quantum Science and Technology (VCQ),
             1090 Vienna, Austria}
\affiliation{Christian Doppler Laboratory for Photonics Quantum Computer,
             Faculty of Physics, University of Vienna, 1090 Vienna, Austria}             

\author{G.~B.~Xavier}
\affiliation{Institutionen för Systemteknik, Linköpings Universitet, 581 83, Linköping, Sweden}

\author{J.~Cari\~{n}e}
\affiliation{Departamento de Ingenier\'{\i}a El\'ectrica,
             Universidad Cat\'olica de la Sant\'{\i}sima Concepci\'on,
             Concepci\'on, Chile}

\author{G.~Saavedra}
\affiliation{Departamento de Ingenier\'{\i}a El\'ectrica,
             Universidad de Concepci\'on, 160-C Concepci\'on, Chile}

\author{E.~S.~G\'omez}
\affiliation{Departamento de F\'{\i}sica, Universidad de Concepci\'on,
             160-C Concepci\'on, Chile}

\author{S.~P.~Walborn}
\affiliation{Departamento de F\'{\i}sica, Universidad de Concepci\'on,
             160-C Concepci\'on, Chile}
\affiliation{Millennium Institute for Research in Optics,
             Universidad de Concepci\'on, 160-C Concepci\'on, Chile}

\author{G.~Lima}
\email{glima@udec.cl}
\affiliation{Departamento de F\'{\i}sica, Universidad de Concepci\'on,
             160-C Concepci\'on, Chile}
\affiliation{Millennium Institute for Research in Optics,
             Universidad de Concepci\'on, 160-C Concepci\'on, Chile}


\begin{abstract}
Quantum key distribution (QKD) provides information-theoretic security
rooted in quantum physics, while high-dimensional (HD) encoding increases
both noise tolerance and secret-key yield. Multicore fibers
(MCFs), a leading platform for next-generation telecom networks, are a
natural substrate for HD-QKD. Field demonstrations over deployed MCFs have so far relied on a hybrid
qudit encoding strategy that combines two path (core modes) with the time-bin
photonic degree of freedom, rather than exploiting the full set of available core modes. Although practical,
this approach incurs intrinsic efficiency penalties that grow with
dimension. Here we implement a four-dimensional ($d=4$) QKD protocol
that directly exploits the full set of core modes of a four-core MCF, operating over an installed MCF network across the
Universidad de Concepci\'on campus under continuous environmental
perturbations. We further benchmark the scheme using superconducting nanowire detectors at
$10\,$dB channel loss, achieving a composable finite-key rate of
$R = 6.19\times 10^{-3}\,$bits/pulse, the highest per-pulse rate
reported to date for HD-QKD at comparable loss. This result establishes core-mode encoding as a viable architecture for realistic, high-rate quantum-secure communications.
\end{abstract}

\maketitle
 
\section{Introduction}
\label{sec:intro}
{%
Quantum key distribution (QKD) is among the most mature quantum
technologies, offering information-theoretically secure key exchange whose
security rests on the laws of quantum mechanics rather than on computational
assumptions~\cite{Bennett1984,Bennett2014,Pirandola2020,Xu2020}.
The looming prospect of cryptographically relevant quantum computers, capable
of breaking widely deployed public-key schemes via Shor's
algorithm~\cite{Shor1994,Shor1997}, has intensified efforts to move QKD out
of the laboratory and into real communication infrastructure. Conventional QKD encodes one bit per photon in a two-dimensional (qubit)
Hilbert space. High-dimensional QKD (HD-QKD), in which each photon constitute a qudit carrying information in a
$d$-dimensional space ($d>2$), offers two interrelated advantages~\cite{Cerf2002,Sheridan2010}:
the tolerable error-rate threshold increases with $d$, improving noise
robustness, and more than one secret bit can in principle be extracted per
detected photon.}

Among the physical platforms for HD-QKD, multicore fibers
(MCFs) are uniquely suited to field operation~\cite{Xavier2020}. An MCF is an optical fiber in which several
independent single-mode cores share a common cladding. MCFs are the
leading platform for space-division multiplexing (SDM)~\cite{richardson10,richardson13}, and are already part of
deployed high-capacity telecom infrastructure \cite{Morita14,Soma18,Hitoshi23,Wu2025,Melo2025,Chen2026}. The same property that makes MCFs
attractive for classical SDM also makes them a natural physical
substrate for HD-QKD: an MCF with $d$ cores supplies a
$d$-dimensional path (core) mode basis, with each core mode corresponding to one
logical state $\ket{k}$. Coherent superpositions of all cores
modes realizes a $d$-dimensional qudit~\cite{Canas2017,Ding2017}. Because all cores share a
common cladding, the differential phase drift between them is orders
of magnitude slower than independent single-mode fibers of
the same length~\cite{DaLio2020}, which helps to keep the
qudit superposition phase-stable across the link. State preparation and measurements are then performed by $d\times d$ multiport beam
splitters, which can be fabricated with low insertion loss directly
in MCFs~\cite{Carine2020}. 

Alternative approaches have combined path with a second encoding
degree of freedom, most commonly time-bin, to build a higher-dimensional
Hilbert space from a smaller number of cores. Following this route, Zahidy~\emph{et al.}~\cite{Zahidy2024}
recently demonstrated a hybrid $d=4$ QKD system on a 52\,km
installed MCF in L'Aquila, using two of the four cores available. The reported composable finite-key rate was 51.5\,kbps at
$22\,$dB of channel loss. Hybrid encodings of this
kind are more robust to perturbations, since fewer core modes have to be
kept coherent simultaneously, but they pay efficiency costs on both
sides of the link. At the transmitter, each qudit spans multiple
pulse slots, so the qudit generation rate is necessarily below the available
laser clock rate. At the receiver, projecting onto time-bin
superposition states requires Franson-like interferometers that discard a
fraction of the detected photons per measurement, compounding with
channel and detector loss~\cite{Alarcon2021}. Pure path encoding eliminates both limitations,
at the cost of requiring simultaneous coherent control of $d$ core modes, a requirement that has so far been met only in laboratory
settings~\cite{Canas2017,Ding2017}.

In this work, a four-dimensional core-mode encoded HD-QKD protocol is
demonstrated over an installed multicore fiber campus network at the
Universidad de Concepci\'on, Chile. The deployed links connect laboratories across separate campus buildings
over distances of approximately 200\,m and 1.3\,km, using permanently
installed fibers that are fully exposed to real-world environmental perturbations
throughout operation. To the authors’ knowledge, this represents the first field deployment of
a core encoded $d=4$ HD-QKD system over a fully installed MCF network under uncontrolled operational conditions,
establishing an important benchmark for realistic high-dimensional
quantum communication. Furthermore, using superconducting nanowire single-photon detectors
(SNSPDs) at 10\,dB of channel loss, the system achieves a composable finite-key rate
of $R = 6.19\times10^{-3}$ bits/pulse, the highest per-pulse finite-key
rate reported to date for HD-QKD at comparable attenuation, surpassing the previous best
time-bin implementation of $3.1\times10^{-3}$ bits/pulse~\cite{Islam2017}
by nearly a factor of two.

\section{Secret Key Rate Estimation}
\label{sec:protocol}

\subsection{Decoy-State Method}
\label{subsec:decoy}

In general, QKD systems use weak coherent pulses (WCPs) rather than ideal
single-photon sources. WCPs follow a Poissonian photon-number distribution, which makes them
vulnerable to photon-number-splitting (PNS) attacks~\cite{PracticalDecoy2005Lo}.
The decoy-state method addresses this vulnerability by requiring Alice to
randomly vary the mean photon number of each pulse among: (i) a signal mean photon number
$\mu$, and (ii) $m$ decoy mean photon numbers with $\nu_1>\ldots>\nu_m$, while $\mu>\nu_1$.
Statistical differences between the gains and error rates observed for these pulses allow one to bound the contribution of true single-photon pulses for secret key generation,
without which the key cannot be proven to be
secure~\cite{PracticalDecoy2005Lo,ConciseSecurity2013Sbinden}.

In the asymptotic limit of infinite pulses sent, the secret-key rate is given by~\cite{PracticalDecoy2005Lo,
ConciseSecurity2013Sbinden}
\begin{align}
R_\infty \geq q \Big[ Q_0 \log_2 d - Q_\mu H_d(E_\mu) f(E_\mu) \nonumber
         \\ + Q_1 \big[ \log_2 d - H_d(e_1) \big] \Big],
\label{eq:Rinf}
\end{align}
where $d$ is the system dimension. $Q_0$ is the vacuum gain, $Q_\mu$ and
$E_\mu$ are the gain and QBER of the signal state, and the
function
\begin{equation}
H_d(x) = -x\log_2\!\frac{x}{d-1} - (1-x)\log_2(1-x),
\end{equation}
is the generalized Shannon entropy for $d$-dimensional systems.
The parameter $f(E_\mu)$ is the error-correction inefficiency, $Q_1$ and
$e_1$ are the gain and error rate of the single-photon pulses, and $q$
is determined by the communication protocol.
For the efficient BB84 protocol used here, $q = p_z^2$, where $p_z$ is the
probability that both Alice and Bob independently choose $\mathcal{Z}$ basis (used exclusively for key generation).

The quantities $Q_0$, $Q_1$, and $e_1$ cannot be measured directly, since
Bob's detectors are not photon-number resolving. Nonetheless, they can be inferred indirectly from experimental data.
In the practical and widely used case of two decoy states (a weak decoy of intensity $\nu$ and a vacuum
decoy), one obtains that $Q_0 = e^{-\mu}Y_0$~\cite{PracticalDecoy2005Lo}, where $Y_0$ denotes the yield
of the vacuum states. A lower bound on the single-photon gain is then given by
\begin{align}
Q_1^L = \frac{\mu^2 e^{-\mu}}{\mu\nu - \nu^2}
\left[ Q_\nu e^\nu - \frac{\nu^2}{\mu^2}Q_\mu e^\mu
      - \frac{\mu^2 - \nu^2}{\mu^2} Y_0 \right].
\label{eq:Q1L}
\end{align}
On the other hand, the upper bound on the single-photon error rate is given by
\begin{align}
e_1^U = \frac{E_\nu Q_\nu \mu e^\nu - \mu e_0 Y_0}{\nu Q_1^L e^\mu},
\label{eq:e1U}
\end{align}
where $e_0 = (d-1)/d$ is the probability of a random dark count in a
detector that is not expected to fire when Alice's and Bob's bases are matched. $Q_\nu$ and $E_\nu$ are the gain and QBER for the decoy state $\nu$.

Substituting these bounds into Eq.~\eqref{eq:Rinf}, one obtains that the lower bound of the secret key rate is given by~\cite{PracticalDecoy2005Lo,
ConciseSecurity2013Sbinden}
\begin{align}
R_\infty \geq q\Big[ Q_0\log_2 d - Q_\mu H_d(E_\mu)f(E_\mu) \nonumber
         \\ + Q_1^L\big[\log_2 d - H_d(e_1^U)\big] \Big].
\label{eq:RinfDecoy}
\end{align}

\subsection{Finite-Key Analysis}
\label{subsec:finite}

The asymptotic rate of Eq.~\eqref{eq:RinfDecoy} assumes an infinite number of
transmitted signals. In any real implementation, only a finite number $N$ of pulses is exchanged,
and the observed quantities $Q_\mu$, $E_\mu$, $Q_\nu$, $E_\nu$ are subject
to statistical fluctuations. Finite-key analysis produces rigorous bounds on these quantities~\cite{ConciseSecurity2013Sbinden,Improve2016Ma}.

Here we follow the approach of Zhang \emph{et al.}~\cite{Improve2016Ma},
adapted to $d$-dimensional systems. A key feature of this method is
that it does not assume a Gaussian profile for the statistical
fluctuations of the observed quantities, which is critical for
high-quality QKD implementations. In the low-QBER regime, the
block-wise QBER distribution becomes strongly asymmetric: the
boundary at zero forces a sharp drop on the low side, observations
of QBER very close to zero become rare, and the distribution's peak
shifts away from its mean, leaving a longer right tail than a
Gaussian would predict. A symmetric Gaussian fit, in this regime,
both overestimates the probability of QBER values close to zero and
underestimates fluctuations toward larger values, corrupting the
finite-key bounds in either direction.

Another interesting property of the method is that it does not rely on the standard Chernoff–Hoeffding aproach \cite{ConciseSecurity2013Sbinden}. Instead, it makes use of the full Chernoff bound when estimating statistical fluctuations. In practice, the usual Chernoff–Hoeffding treatment simplifies this bound into a convenient exponential form, which introduces additional looseness in the resulting estimates. By retaining the full expression, the method captures more accurately the relation between the observed counts and the underlying probabilities, leading to tighter confidence intervals for the estimated parameters.

According to the two-decoy finite-key method of Zhang
\emph{et al.}~\cite{Improve2016Ma}, with Chernoff-based inverse bounds and
random sampling, a secret key of length $K$ that is
$\epsilon_{\sec}$-secure and $\epsilon_{\mathrm{cor}}$-correct
satisfies~\cite{Improve2016Ma,ConciseSecurity2013Sbinden}
\begin{align}
K \geq \Big[ & M_0^{zL}\log_2 d - M_\mu H_d(E_\mu)f(E_\mu) \nonumber\\
             & + M_1^{zL}\big(\log_2 d - H_d(e_1^U)\big)
               - 6\log_2\!\tfrac{21}{\epsilon_{\sec}}
               - \log_2\!\tfrac{2}{\epsilon_{\mathrm{cor}}} \Big],
\label{eq:finiteK}
\end{align} where $R = K/N$, and $M_a$ is the number of detections from pulses of type $a$
(for signal: $a=\mu$; weak decoy: $a=\nu$; vacuum: $a=0$). $\epsilon_{\mathrm{cor}}$ quantifies the probability that Alice and Bob obtain different final secret keys after error correction and verification, while $\epsilon_{\mathrm{sec}}$ quantifies Eve's possible information about the final key. A protocol is considered secure when both probabilities are sufficiently small. Next, we describe the method adapted for $d$-dimensional systems such that the interested reader can reproduce it for future use. Nonetheless, we leave the complete mathematical derivation on Appendix~\ref{app:chernoff}. 

First, inverse Chernoff bounds Eq.~(\ref{eq: A1}) are applied to the experimentally observed
detection counts associated with each WCP of the protocol. This provides confidence intervals for the expected gains
$\mathbb{E}[Q_a] = \mathbb{E}[M_a]/N_a$ of the signal, weak-decoy, and vacuum
WCPs, where $N_a$ is the number of transmitted pulses of type $a$. The resulting confidence intervals for the gains are then used in the
standard decoy-state relations to derive a lower bound on the number of
only single-photon detections in the $\mathcal{Z}$ basis across all WCPs,
denoted by $M_1^{*zL}$, as shown in Eq.~(\ref{eq:M1starL}).

The expected number of single-photon detections across all WCPs is
then converted through Eq.~(\ref{eq:Msz1L}) into a lower bound on the number
of single-photon events originating only from the signal WCPs, denoted
by $M_1^{zL}$. This step uses a symmetric Chernoff Eq.~(\ref{eq:simcher}) bound to
account for the statistical fluctuation between the expected and observed
number of such events. An analogous procedure is applied to the vacuum contributions in order to
obtain a lower bound $M_0^{zL}$ on the number of vacuum detections in the
$\mathcal{Z}$ basis associated with signal WCPs.

Finally, the phase-error rate of single-photon events is estimated from the
error statistics ($E_aM_a$) measured in the $\mathcal{X}$ basis. Specifically, the upper limit $e_1^{xU}$ is obtained using the Chernoff bounds (see Eq.~(\ref{eq:e1ux})). A random sampling argument is then applied to convert this value into an upper bound $e_1^{U}$ on the phase-error rate in
the $\mathcal{Z}$ basis, as shown in Eq.~(\ref{eq:randomsampling}).

The resulting bounds $M_1^{zL}$, $M_0^{zL}$, and $e_1^{U}$ are substituted
into Eq.~(\ref{eq:finiteK}) to obtain the composable secret key length, with the security and correctness parameters $\epsilon_{\mathrm{sec}} =3.80 \times 10^{-8} $ and
$\epsilon_{\mathrm{cor}} = 1.00 \times 10^{-15}$.

\section{Experimental Setup}
\label{sec:setup}

\begin{figure*}[htbp]
    \centering
    \includegraphics[width=\linewidth]{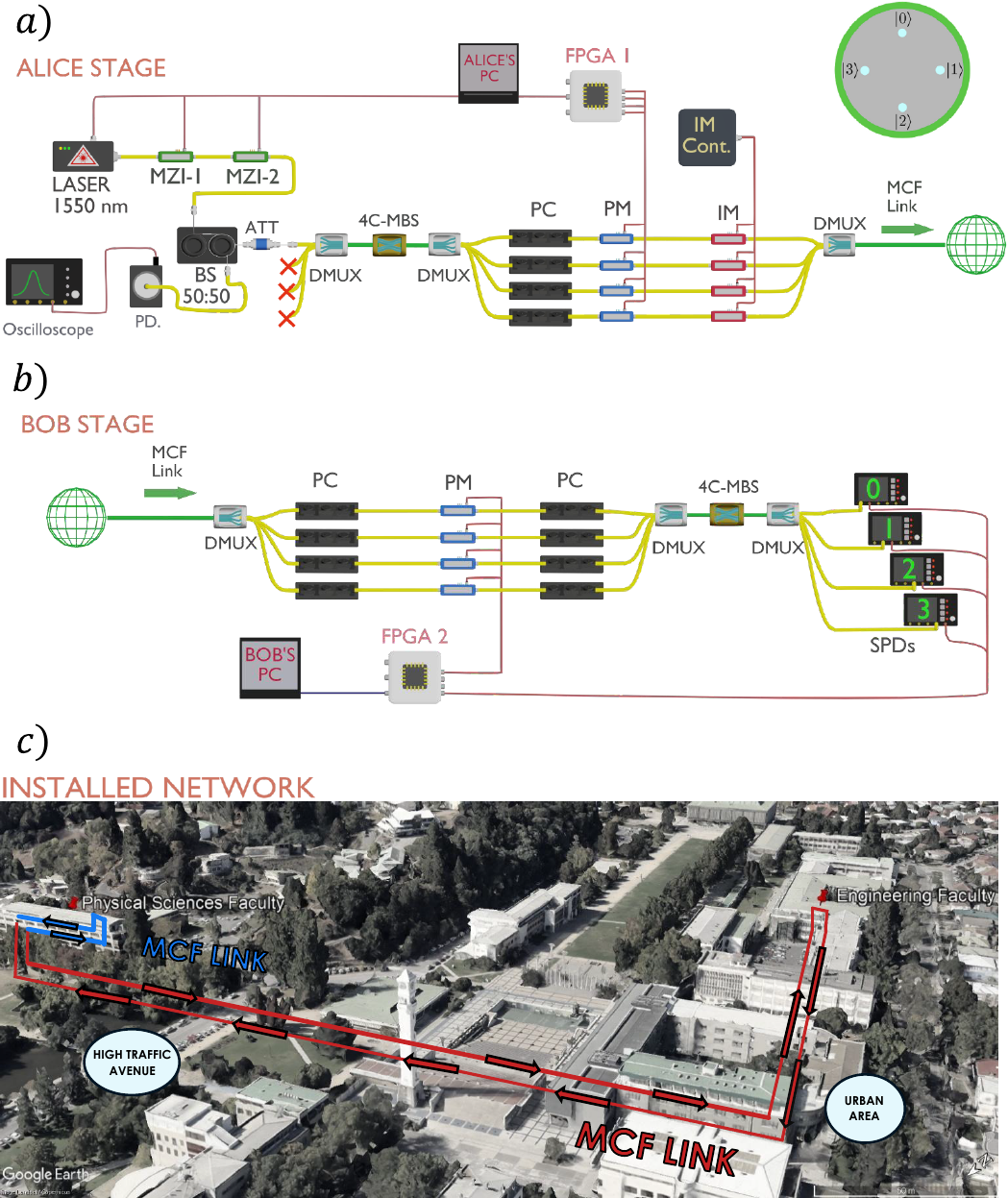}
    \caption{Schematic of the core-encoded four-dimensional QKD system and network.
    Alice (a) prepares four-dimensional qudit states; Bob (b)
    measures them in two different bases. (c) The states are transmitted through different links of an MCF network installed over the campus of the Universidad de Concepci\'{o}n, Chile.
    PC: Polarization Controller; PM: Phase Modulator; IM: Intensity
    Modulator; Att: Optical Attenuator; MCF: Multicore Fiber;
    MUX/DMUX: MCF Multiplexer/Demultiplexer; 4C-MBS: Four-Core Multiport
    Beam Splitter; SPD: Single-Photon Detector.}
    \label{fig:setup}
\end{figure*}

The HD-QKD system consists of a prepare-and-measure apparatus, where transmitter and receiver are connected through a field-deployed MCF network as illustrated
in Fig.~\ref{fig:setup}. Four-dimensional quantum states (ququarts, $d=4$) are encoded in the
core modes of a four-core MCF: the logical basis states $\{\ket{0},\ket{1},\ket{2},\ket{3}\}$ are associated with the four cores (see the inset of Fig.~\ref{fig:setup}(a)).
The setup is a large fiber interferometer based on different MCF technologies. Superpositions of the logical states are created using a four-core MCF multiport beam splitter (4C-MBS) at Alice. These are then transmitted through the installed MCF network to Bob, where the four modes interfere at a
second 4C-MBS. The HD-QKD protocol is implemented using the phase-coding scheme, where Alice controls the initial phase between the core modes and Bob the final ones~\cite{Pirandola2020,Xu2020}. In the following sub-sections we describe each stage of the system in details.

\subsection{Alice: State Preparation}
\label{subsec:alice}

Alice's preparation stage, controlled by a field-programmable gate array
(FPGA) eletronics, begins with a continuous-wave telecom laser at 1550\,nm (see Fig.~\ref{fig:setup}(a)).
The laser output is carved into 5 ns pulses at a repetition rate of 2\,MHz by
two cascaded Mach--Zehnder interferometers (MZIs) driven by the FPGA.
A tap coupler diverts a small fraction of the output to a photodetector
and oscilloscope for real-time power and pulse monitoring. The main arm proceeds to
an optical attenuator (Att) that fine adjust the pulse intensity. Phase-drifts and different voltage levels are then used on MZI-2 to properly generate the 
WCPs required on the two-decoy QKD method.

The attenuated pulses enter an initial MCF through a spatial demultiplexer (DMUX), which is a commercial fan-in/fan-out
device that maps each core of a MCF to an individual single-mode fiber. The pulses are then sent to a 4C-MBS ~\cite{Carine2020}, which splits the input equally
across all four cores. The output cores of the 4C-MBS are coupled back to four single mode fibers (SMFs) using another DMUX in the reverse order. The SMFs are connected to a set of four phase modulators (PMs), one for each SMF. With the 4C-MBS and PMs, Alice can create an equal-amplitude core-encoded qudit superposition. The entire single-mode fiber section is only $\simeq 5$~m long, and is covered on isolation materials such that their contribution to the interferometer's drifts is negligible. The generated qudits are then given by
\begin{equation}
\ket{\Psi} = \frac{1}{2}\bigl(
  e^{i\phi_0^A}\ket{0} + e^{i\phi_1^A}\ket{1}
+ e^{i\phi_2^A}\ket{2} + e^{i\phi_3^A}\ket{3}\bigr),
\label{eq:Alicestate}
\end{equation}
where $\phi_k^A$ is the phase applied by the PM acting on core $k\in\{0,1,2,3\}$.
By choosing the set of phases $\{\phi_k^A\}$ properly, Alice prepares
one of eight qudit states (given in Fig.~\ref{fig:stateprep} in Appendix~\ref{app:d4BB84}.),
corresponding to one of two mutually unbiased bases $\mathcal{Z}$ or
$\mathcal{X}$, as required by the HD-BB84 protocol. By design, neither $\mathcal{Z}$ nor $\mathcal{X}$ coincides with the core (logical) basis. Rather,  both are composed of superposition states, such that fast PMs can toggle between them. In each path a polarization controller (PC) aligns the optical field to the
acceptance axis of each PM. The four PMs are independently driven by the FPGA to apply individual
phases to each core. Intensity modulators (IM) in each arm are used for aligning purposes before each QKD session. Finally, a DMUX couples the four paths back into MCFs for transmission
over the installed MCF network.

\subsection{Bob: Measurement System}
\label{subsec:bob}

At Bob's station, a DMUX is used to separate light from the four cores into another isolated $\sim 5$m-long section of independent SMFs, which are each connected to a PC, PM and second PC (see Fig.~\ref{fig:setup}(b)). The first set of PCs aligns the polarization in each SMF to the acceptance
axis of the corresponding PM. The PMs then apply one of two fixed phase patterns to select the measurement
basis: $\vec{\Phi}_{\mathcal{Z}}^B=\{0,0,0,0\}$ for the $\mathcal{Z}$ basis, or
$\vec{\Phi}_{\mathcal{X}}^B=\{\pi,0,0,0\}$ for the $\mathcal{X}$ basis.  
The second set of PCs finely tunes the polarization to ensure mode
indistinguishability at the 4C-MBS. The four paths are routed through compact DMUX stages before and after
the 4C-MBS, likewise Alice's stage. Each output port of the 4C-MBS is monitored by a single-photon detector (SPD) to conclude the measurement procedure.

The resulting measurement basis states are explicitly given by~\cite{Carine2020}
\begin{eqnarray}
|\psi_{0}\rangle= \frac{1}{2}(e^{i\phi_{0}^{B}} |0 \rangle+e^{i\phi_{1}^{B}} |1 \rangle+e^{i\phi_{2}^{B}} |2 \rangle+e^{i\phi_{3}^{B}} |3 \rangle) , \nonumber \\
|\psi_{1}\rangle= \frac{1}{2}(e^{i\phi_{0}^{B}} |0 \rangle+  e^{i\phi_{1}^{B}} |1 \rangle- e^{i\phi_{2}^{B}} |2 \rangle- e^{i\phi_{3}^{B}} |3 \rangle), \nonumber \\
|\psi_{2}\rangle= \frac{1}{2}(e^{i\phi_{0}^{B}} |0 \rangle- e^{i\phi_{1}^{B}} |1 \rangle+e^{i\phi_{2}^{B}} |2 \rangle-e^{i\phi_{3}^{B}} |3 \rangle), \nonumber \\
|\psi_{3}\rangle= \frac{1}{2}(e^{i\phi_{0}^{B}} |0 \rangle-  e^{i\phi_{1}^{B}} |1 \rangle- e^{i\phi_{2}^{B}} |2 \rangle+ e^{i\phi_{3}^{B}} |3 \rangle),
\label{eq:Bobstates}
\end{eqnarray} where $\phi_k^{B}$ is the phase applied by the second modulator in the core mode $k$. Projection onto state $|\psi_{k}\rangle$ is associated to the SPD in core mode $k$. 

For the QKD experiment, two detector technologies were employed. We first used InGaAs avalanche photodiode SPDs to characterize the network links and to test the QKD system operating with this low-cost technology, demonstrating that the system remains viable with more commercially accessible hardware. In this case, Bob's mean detection efficiency is limited to $\sim 7.5\%$. We then used modern SNSPDs, with mean detection efficiency of $\sim 85\%$, to define the practical performance bound attainable by the system, setting its finite-key generation rate benchmark. Bob's FPGA synchronizes the detection gate windows with Alice's pulsed laser and controls Bob's PMs.

\subsection{MCF Network}
\label{subsec:channel}

The QKD system was tested on a multi-core fiber network installed across the campus of the
Universidad de Concepci\'on.
The network employs custom-fabricated cables, each one containing two
Fibercore~SM-4C1500 four-core homogeneous MCFs. Each multicore fiber has four identical cores of diameter 8\,\textmu m arranged in a
square geometry with a center-to-center pitch of 50\,\textmu m (nominal)
within a cladding of $124\pm1$\,\textmu m diameter.
The mode field diameter at 1550\,nm is 7.4--8.5\,\textmu m and the operating wavelength range is 1520--1650\,nm.
The fiber coating is dual acrylate with a diameter of $245\pm12$\,\textmu m
and an operating temperature range of $-55$ to $+85\,^{\circ}$C.
Each core supports single-mode propagation in the telecom C and L bands,
with a nominal attenuation of 0.2\,dB/km.

The campus network is arranged in a star topology, with a central laboratory serving as the hub that interconnects three peripheral laboratories (see Fig.~\ref{fig:setup}(c)) \cite{Melo2025}. The main laboratory, which houses Alice's and Bob's apparatus, is located on the second floor of the Faculty of Physical and Mathematical Sciences. The hub resides on the same floor of the building, separated from the main laboratory by several tens of meters of MCF that constitute the first deployed link. From the hub, a vertical link rises through the building to a third laboratory on the sixth floor, while a second link extends across the university campus to reach a fourth laboratory in the Faculty of Engineering. Together, these three fiber spans form the physical backbone of the network and define the set of point-to-point channels available to the QKD system. Three distinct operating round-trip configurations are implemented over this infrastructure and are referenced throughout the paper by

\begin{itemize}
  \item \textbf{BtB} (back-to-back): Alice is connected directly to Bob through
    three 5\,m patchcords (15\,m of total fiber), entirely within
    the main laboratory. This configuration serves as the reference baseline.
  \item \textbf{P6} (sixth-floor link): Alice's output is routed from the second
    floor through the internal building infrastructure to the laboratory on the
    sixth floor and back, yielding a round-trip length of approximately 200\,m.
  \item \textbf{PtE} (Physics-to-Engineering link): Alice's output is routed out
    of the main building, through the underground fiber infrastructure that
    crosses the campus plaza, into the laboratory at the Faculty of Engineering,
    and back, yielding a round-trip length of approximately 1300\,m.
\end{itemize}

The round-trip configuration of each link is realized over a pair of 4-core MCFs:
one MCF carries the encoded states from Alice toward the remote laboratory, while
the second returns them to the main laboratory, where Bob's measurement apparatus
is located. Further details of these links are given in Table~\ref{tab:channels}.

\begin{table}[t]
\centering
\begin{tabular}{lccc}
\toprule
Link & $\eta_L$ (dB) & $L$ (m) & Eff.\ MCF (km) \\
\midrule
BtB & 2.16  & 15   & 10.8 \\
P6  & 9.36  & 200  & 46.8 \\
PtE & 10.70 & 1300 & 53.5 \\
\bottomrule
\end{tabular}
\caption{Channel configurations of the deployed network. For each link we
report the total measured attenuation $\eta_L$ (averaged over the four MCF
cores) and the round-trip optical length $L$. The last column gives the
equivalent length of MCF (assuming 0.2\,dB/km) that would yield
the same attenuation in the absence of connectors and splices.}
\label{tab:channels}
\end{table}

As is typical of testbed networks, where the physical configuration is frequently reconfigured and reconnected, the channel loss is dominated by fiber-to-fiber connections and splices rather than by intrinsic fiber propagation. For the PtE link, the bare fiber contributes only $\approx0.13$\,dB of the
total 10.70\,dB; the remaining $\approx10.6$\,dB arises from stress-related core misalignment from butt-coupled MCF connectors, fusion splices, and DMUXs. As a result, $\eta_L$ is essentially uncorrelated with $L$: the P6 link (200\,m) and PtE link (1300\,m) have nearly the same total attenuation. Nonetheless, the PtE cable traverses a road, a parking lot, and a pedestrian plaza
with substantial daytime activity, producing a noisier and more variable differential-phase environment than the P6 link.

\section{Measurement procedure and Network characterization}

\subsection{Polariz ation stability}

\begin{figure}
  \centering
  \includegraphics[width=\linewidth]{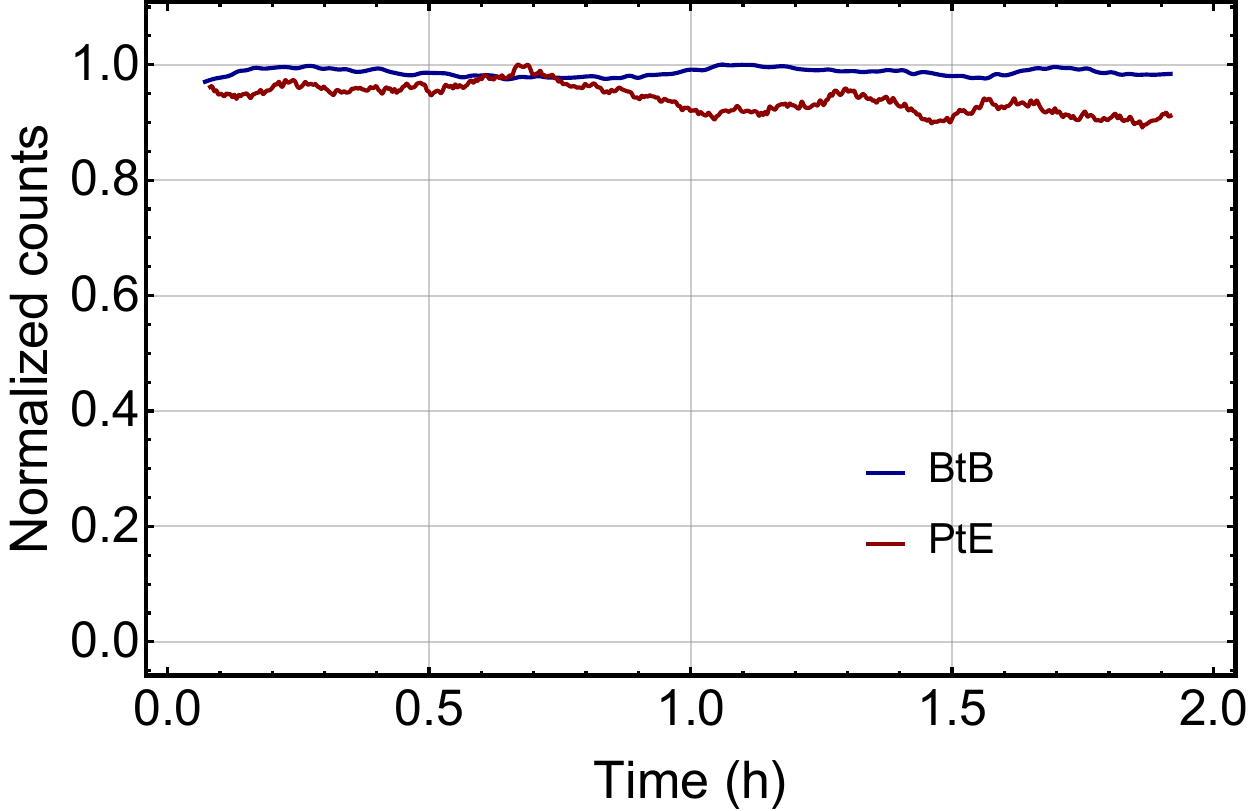}
  \caption{Polarization stability of the deployed links. Normalized single-photon count rate, integrated in 100ms bins, recorded with a fixed polarization analyzer for the back-to-back (BtB) configuration and the installed Physics-to-Engineering (PtE) link. After an initial manual alignment, no active polarization control is applied. The count rate remains stable over more than two hours in both settings, demonstrating that environmental drifts in the laboratory and across the deployed campus link are slow compared with the duration of a typical QKD session.}
  \label{fig:pol_stability}
\end{figure}

Polarization stability of the installed links is critical. Independent polarization evolution along the different cores of the MCF would compromise the coherent propagation of the core modes used by the QKD protocol. To prevent this, the polarization state of each core is manually aligned prior to every QKD run. At Alice's station, the PCs are adjusted to match the acceptance axis of her PMs. At Bob's station, the first set of PCs is aligned with the acceptance axis of his PMs, while the second set is tuned for polarization-mode indistinguishability at the 4C-MBS. Once this initial alignment is completed, the PCs remain static and no active polarization stabilization is employed such that slow environmental perturbations are allowed to act freely on the channel. Representative polarization-stability measurements for the BtB and PtE links are shown in Fig.~\ref{fig:pol_stability}. In both cases, the polarization state remains essentially constant for more than two hours, confirming that the in-laboratory and outside-plant environments are sufficiently stable on this timescale. This window comfortably exceeds the time required to initialize and complete the QKD sessions reported below, eliminating the need for active polarization control. 

\subsection{Phase Stabilization and Data Acquisition}
\label{subsec:calibration}

\begin{figure}
  \centering
  \includegraphics[width=0.9\linewidth]{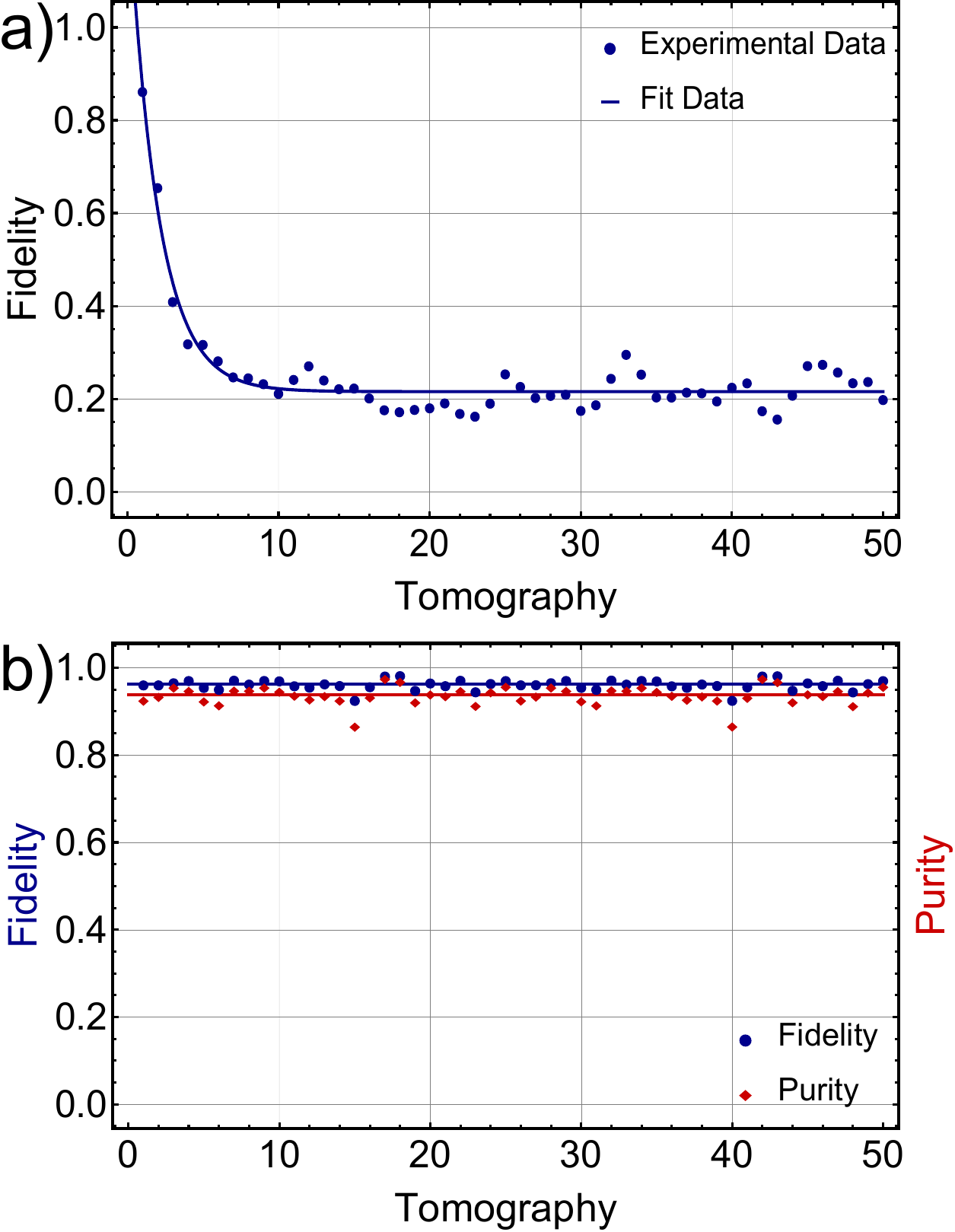}
  \caption{Characterization of the PtE MCF link. (a) Free evolution of the transmitted state over $\sim 20\,\mathrm{s}$, reconstructed via sequential quantum state tomography (for the first input state defined in Fig.~\ref{fig:stateprep}). Each projector is integrated for $100\,\mathrm{ms}$, and a full tomographic reconstruction is completed in $\sim 400\,\mathrm{ms}$. The slow phase drifts induced by thermal and mechanical perturbations can be seen as unitary rotations of the prepared state, progressively degrading its overlap with the target. (b) Tomographic reconstructions acquired within a single $100\,\mathrm{ms}$ interval following the preparation of the same input state, repeated over $50$ independent realizations. The reconstructed states exhibit a mean fidelity of $\bar{F}=0.96\pm0.01$ with respect to the target one and a mean purity of $\bar{P}=0.94 \pm 0.02$, demonstrating that the residual stochastic phases $\delta\phi_k(t)$ remain stationary over $100\,\mathrm{ms}$. Thus, $100\,\mathrm{ms}$ is an operational acquisition window, which was then employed in the QKD protocol.}
  \label{fig:tomo}
\end{figure}

In our phase-coding QKD scheme, information is encoded in the relative
phases across the four core modes. Random phase fluctuations between different cores
must therefore be compensated before each key-generation window.
Although the common cladding of the MCF suppresses differential phase
drift between cores~\cite{DaLio2020}, residual stochastic phases
$\delta\phi_k(t)$ remain, driven primarily by temperature and mechanical
perturbations along the link. This can be seen in Fig.~\ref{fig:tomo}(a) where we prepare the first state given in Fig.~\ref{fig:stateprep}, and let the PtE MCF link evolve freely while monitoring the photon state evolution using quantum tomography \cite{Daniel2001,Daniel2009}. The integration time for recording the single counts is 100\,ms, and each tomography takes $\sim400\,$ms to be concluded. Thus, Fig.~\ref{fig:tomo}(a) spans a total time of $20$\,s of channel evolution. The slow phase drifts can be seen as random rotations on the states prepared by Alice, which leads to errors in Bob's measurement outcomes and thus an increased QBER in a QKD session. Importantly, we find that these phases are stable over periods of at least $100$\,ms, which is the key property that makes a 100\,ms acquisition window viable. This can be seen in Fig.~\ref{fig:tomo}(b), where the tomographies are now performed during the time window of 100\,ms after the same Alice's state has been prepared. A procedure that has been repeated 50 times. In this case, high overall fidelities and purities are obtained for the reconstructed states with a mean value of $\bar{F}=0.96 \pm 0.01$ and $\bar{P}=0.94 \pm 0.02$, respectively. 

Once a viable operational detection time is identified, it is necessary to ensure that Alice's states and Bob's measurements remain properly implemented over time despite slow phase fluctuations along the MCF. For this, we employ a calibration procedure based on Alice's phase modulators. Prior to each measurement session, the modulators are biased so that, at the input of Bob's 4C-MBS, the effective relative phases among the four cores are nulled up to a global phase. Specifically, assuming that $\delta\phi_k(t)$ represents the stochastic phase accumulated on
core~$k$ during propagation, and by denoting $\alpha_k$ as the compensation phase
applied by Alice's modulator on that core, the condition for perfect
constructive interference at detector~$D_0$, corresponding to core mode~0, is
\begin{equation}
{%
\bigl[\alpha_k + \delta\phi_k(t)\bigr]
- \bigl[\alpha_0 + \delta\phi_0(t)\bigr]
\equiv 0 \pmod{2\pi}.
}
\label{eq:phase-balance}
\end{equation}
Note that $\alpha_0$ does not need to be zero: the equation constrains only the
\emph{phase differences} between cores, leaving a free global phase $\alpha_0$
that has no physical consequence for the interference. So, in our experiment we use a perturb-and-observe algorithm to modify Alice's PM phases, presented in detail at \cite{Carine2024}, 
which finds the set $\{\alpha_k\}$ satisfying Eq.~\eqref{eq:phase-balance}, and that maximizes the count rate at $D_0$ when sending the state 
$\{0,0,0,0\}$. The fraction $F_{e}$ of photons detected at $D_{j\neq 0}$, thus provides a direct,
real-time indicator of phase imperfection in the channel.  We define the threshold $T$ as the maximum allowable for this erroneous count rate. Thus, phase calibration is performed until  $F_{e} \leq T$, at which point the system begins the protocol procedure (Run phase). During the Run phase, Alice's total phase on core~$k$ is
$\alpha_k + \phi_k^{A}$, where $\phi_k^{A} \in \{0,\pi\}$ is one of the phases required to encode a chosen QKD state. The compensation phases $\{\alpha_k\}$ are held fixed throughout the
Run window and updated again only in the next Adjust phase. After each Run window (of $100$\,ms), the system checks the relation $F_{e} \leq T$, if satisfied, another Run phase is performed. If not, the system returns to adjust the phases. Figure~\ref{fig:tomo}(b) constitutes also an example of the resulting alignment, where a QKD protocol state is prepared and measured over time while considering $T=5\%$.

\section{QKD Results and Discussion}
\label{sec:results}

\subsection{QBER Distributions and Threshold Analysis}

\begin{figure}
  \centering
\includegraphics[width=\linewidth]{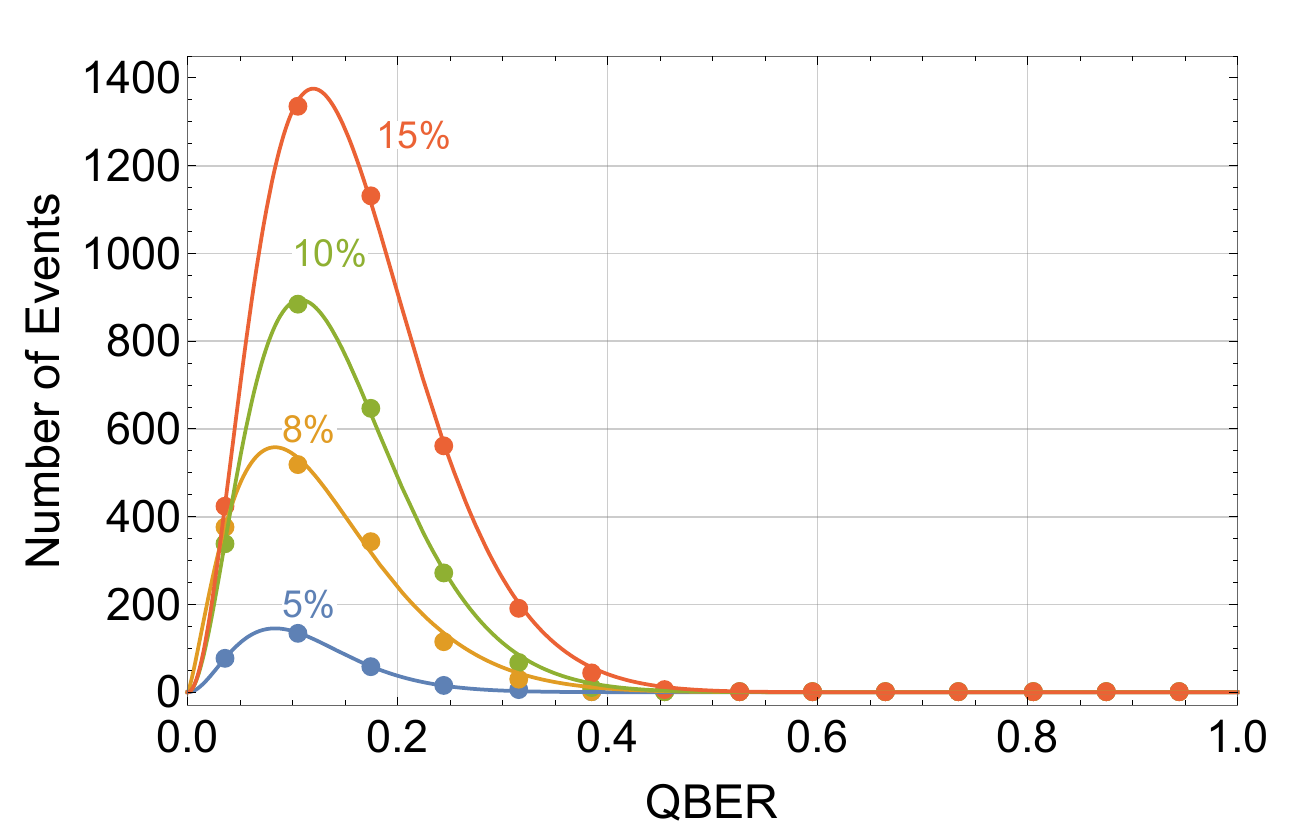}
\caption{Distributions of the QBER measured in $100\,\mathrm{ms}$ windows on the PtE link, for threshold values $T$ ranging from $5\%$ to $15\%$. Each histogram is fit with a Beta distribution (solid lines), which captures the asymmetric, bounded character of the data: the hard boundary at zero suppresses the low side, the peak is shifted away from the mean, and a longer right tail extends toward larger errors. As $T$ is tightened, the distribution narrows and shifts toward $\mathrm{QBER}\approx 0$, at the cost of a reduced duty cycle (fewer accepted windows).}
\label{fig:fidelity-thresholds}
\end{figure}

We initially investigated the statistical distribution of the QBER and its dependence on the threshold $T$, in order to properly characterize optimal operation conditions for secret key generation. Figure~\ref{fig:fidelity-thresholds} shows the QBER distributions for the PtE link as a function of the threshold $T$, varied from $5\%$ to $15\%$. We focus on the PtE link because it represents the more demanding scenario: its longer cross-campus route is subject to higher environmental noise, resulting in a broader QBER distribution than the P6 loop under comparable thresholds. The histograms display the asymmetric, bounded character anticipated previously: the hard boundary at zero suppresses the low side, the peak is shifted away from the mean, and a longer right tail extends toward larger errors. This asymmetry is most pronounced at tight thresholds, where the distribution is pushed against the lower boundary, but remains visible across the entire range of $T$ explored. A symmetric Gaussian fit is manifestly inadequate in this regime, as it would overestimate the probability of near-zero QBER and underestimate the upper-tail fluctuations. This justifies the use of the Zhang \emph{et al.} analysis~\cite{Improve2016Ma}, which makes no Gaussian assumption on the statistics of the observed quantities. 

The dependence of the distribution on $T$ illustrates the trade-off inherent to the phase-stabilization algorithm: a tighter threshold selects lower-error windows but increases the re-adjustment frequency, thereby reducing the duty cycle ($\mathcal{D}$). The net effect on the total secret key length depends on the interplay between $\mathcal{D}$ and $T$, and the optimal threshold is both link- and detector-specific. The parameter that determines whether a positive key rate is achievable at all, and therefore sets the maximum operating distance, is the mean QBER. At the security boundary ($\mathrm{QBER} = 18.75\%$ for $d=4$) the key rate vanishes regardless of how many windows are accepted. Thus, reducing the QBER is therefore the primary mechanism for extending range.

\subsection{Secret Key Rates over the MCF Network}
\label{subsec:keyrate}

This work reports two main results, one in this sub-section and one in the next. The first is a demonstration of a four-dimensional QKD session based on pure core-mode encoding, implemented for the first time over an installed MCF network under uncontrolled real-world perturbations. Previous deployed-fiber demonstrations have instead relied on hybrid encodings that combine path with an auxiliary photonic degree of freedom, not fully exploiting all the cores available in the MCF link. This establishes core-mode based HD-QKD as viable under field conditions and sets a quantitative reference for the design of future MCF-based quantum networks.

The second is a composable finite-key rate of $6.19\times10^{-3}$\,bits/pulse, obtained with SNSPDs at 10\,dB of channel loss. This is the highest per-pulse rate reported to date for HD-QKD at comparable attenuation, nearly a two-fold improvement over the previous best. Beyond this record figure, the result shows that hybrid encodings can be avoided without compromising the secret-key rate.

\begin{figure}[t]
  \centering
\includegraphics[width=0.95\linewidth]{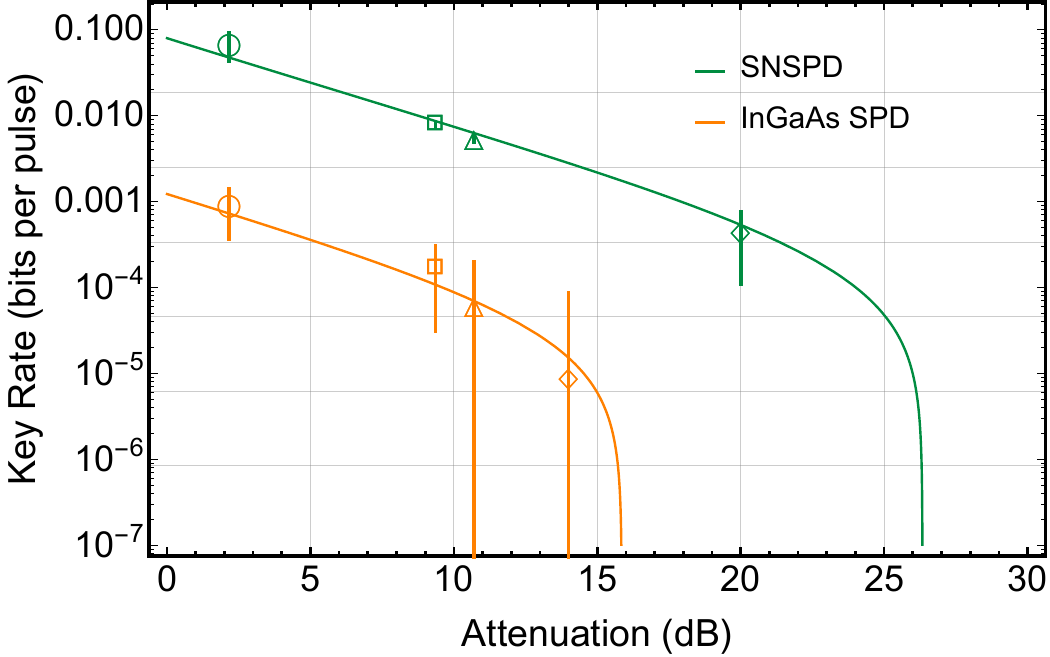}
  \caption{Asymptotic key rate (bits/pulse) vs.\ channel attenuation
  while using InGaAs SPDs and SNSPDs.
  Points: recorded experimental rates. Curve: asymptotic model. ($\circ$) represents the results for the BtB link; ($\Box$) represents the results for the P6 link; ($\triangle$) represents the results for PtE link and; ($\diamondsuit$) represents the results for channels with extra attenuation to test the viability for secret key generation with higher losses. The key generation values are given explicitly in Appendix~\ref{app:data}.}
  \label{fig:keyrate}
\end{figure}

Figure~\ref{fig:keyrate} summarizes asymptotic secret key rates obtained with InGaAS SPD operating at $T=5\%$, and for SNSPDs operating with $T=3\%$  (corresponding to the lowest $T$ achievable of each technology). Like we said, two detector technologies were employed: InGaAs avalanche photodiode SPDs, to assess system viability with commercially accessible hardware, and SNSPDs, to establish the practical performance bound of the system. All the experimental data recorded and used to calculate such secret key rates are explicitly given in Appendix~\ref{app:data}.

For the InGaAS SPDs, lower key rates are obtained as expected. Nonetheless, since we are interested on testing the limits of the system, it is interesting to see that a more commercial version of this QKD system can operate up to $14$ dB of channel losses. In the ideal case, this is equivalent to the attenuation obtained over $\sim70$ kms of modern multicore fibers. In this investigations we worked with a clock rate set to 2~MHz because: (i) this already allows one to determine with good precision the most relevant parameter of a QKD system, which arguably is the secret key rate per pulse transmitted; and (ii) it avoids saturation of the SNSPDs at low attenuation regimes. Nonetheless, all the active optical devices can operate with faster electronics. Values up to 500~MHz of clock rate can be easily adopted. For the experimental data at $14$ dB of channel losses, we obtain a secret key rate per pulse of $8.62\times 10^{-6} \pm 7.63\times 10^{-5}$. Thus, the system with a viable commercial technology would be able to reach in this case a low but not negligible final asymptotic key rate $>1$~kbps. 

\begin{table}[t]
\centering
\begin{tabular}{lc}
\toprule
Quantity & Value \\ \midrule
Loss (dB) & 10 \\
$N$ & $6.130\times10^9$ \\
$M_z$ & $8.18\times10^7$ \\
$M_{\mu z}$ & $80\,904\,263$ \\
$M_{\nu z}$ & $904\,329$ \\
$M_{\mu x}$ & $677\,165$ \\
$M_{\nu x}$ & $7\,640$ \\
$M_{0 z}$ & $2\,082$ \\
$E_{\mu z}M_{\mu z}$ & $1\,703\,281$ \\
$E_{\nu x}M_{\nu x}$ & $252$ \\
$E_{0 x}M_{0 x}$ & $13$ \\
$e_{\mathrm{det}}^z$ & $1.98\%$ \\
$e_{\mathrm{det}}^x$ & $2.85\%$ \\
\midrule
$M_0^{zL}$ & $ 54\,842$ \\
$M_1^{zL}$ & $35\,290\,908$ \\
$e_1^U$ & $7.29\%$ \\
$f(E_\mu)$ & $1.05$ \\
\midrule
$\epsilon$                 & $10^{-10}$ \\     
$\epsilon_{cor}$           & $10^{-15}$ \\
$\epsilon_{sec}$           & $3.8\times10^{-8}$ \\
\midrule
$K$ & $3.80\times10^7$ \\
$R$ & $6.19\times10^{-3}$ \\
\bottomrule
\end{tabular}
\caption{Experimental data recorded for the finite-key rate estimation.
$N$: total pulses emitted; $M_z$: $\mathcal{Z}$-basis detections;
$e_{\mathrm{det}}^z$, $e_{\mathrm{det}}^x$: average QBERs in
$\mathcal{Z}$ and $\mathcal{X}$;
$M_0^{zL}$, $M_1^{zL}$: lower bounds on vacuum and single-photon detections;
$e_1^U$: upper bound on single-photon error rate;
$f(E_\mu)$: error-correction inefficiency;
$K$: secret key length; $R$: secret-key fraction.}
\label{tab:finitekey}
\end{table}

The asymptotic green curve modeling the results obtained with SNSPDs, define the secret key rate limits for our core-encoding QKD system. For instance, much higher losses are tolerated by the QKD system such that it could reach, in the ideal case, MCF propagation distances greater than 100~Kms. This is shown in Fig.~\ref{fig:keyrate}, where we have been able to achieve positive secret key generation rates at 20 dB of artificial channel attenuation. The impact of detector technology on the core-encoding QKD system can be directly assessed over the deployed PtE link, for instance. With InGaAs SPDs, we measured a secret key rate of $R_\infty = 6.04 \times 10^{-5}$ bits per pulse, whereas SNSPDs yielded $R_\infty = 5.36 \times 10^{-3}$ bits per pulse, corresponding to an improvement of two orders of magnitude.

\subsection{Finite-Key rate benchmark}
\label{subsec:snspd}

\begin{figure}[t]
  \centering
  \includegraphics[width=0.95\columnwidth]{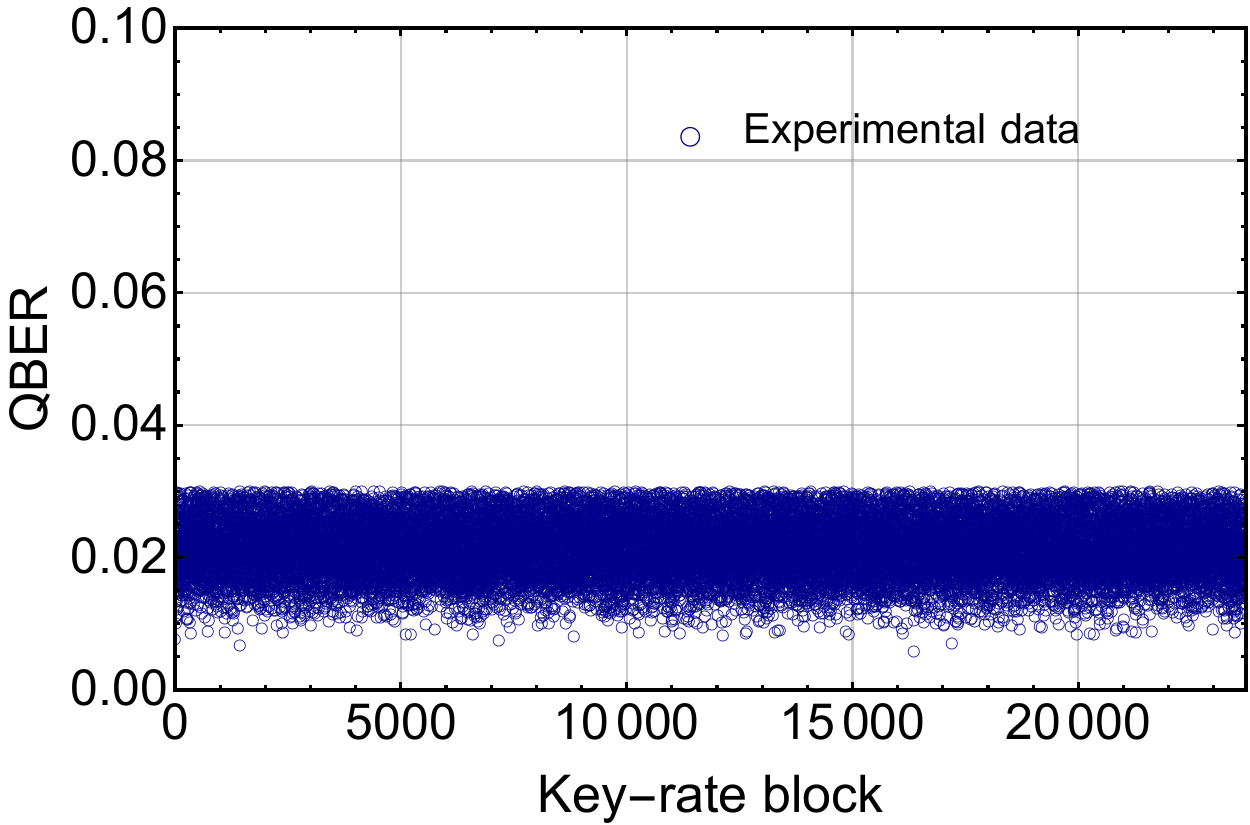}
  \caption{QBER distribution, integrated over blocks of 100ms, for the entire $M_z=8.18\times10^7$ detections registered in the QKD session. The distribution center remains mainly within $1.5\%$--$2.5\%$, with weak
  dependence on block counts, indicating that the spread is dominated
  by detector dark counts and timing jitter rather than phase drifts.}
  \label{fig:qberfinite}
\end{figure}

Now we study the core-encoded QKD performance while considering finite statistics in the key generation procedure. The asymptotic analysis based on the two-decoy method is relevant to define upper bounds on the QKD performance of the system. Nonetheless, it assumes an ideal situation where an infinite number of WCPs are sent to Bob, such that probabilities can be measured directly. In practice. however, only a finite number of WCPs can be sent, and the implemented measurements are subject to statistical fluctuations that must properly be modelled to guarantee the security of the generated key. Another interesting aspect of measuring the composable finite-key generation rate for the core-encoded QKD scheme, is that this value can be used to directly compare our system against several modern QKD experiments that also rely on finite-key analysis. 

\begin{table*}[t]
\centering
\begin{tabular}{lcccl}
\toprule
Reference &Dimension &Encoding & Loss (dB) & Rate (bits/pulse) \\
\midrule
Zahidy \emph{et al.}~\cite{Zahidy2024}
  & 4 & core+time-bin & $10$ & $1.4\times10^{-3}$ (finite)$^a$ \\
  Islam \emph{et al.}~\cite{Islam2017}
  & 4 & time-bin & $10$ & $3.1\times10^{-3}$ (finite)$^a$  \\ 
  W. Li \emph{et al.}~\cite{Li2023}
  & 2 & polarization & $10$ & $8.9\times10^{-3}$ (finite)$^a$  \\
  This work
  & 4 & core modes & 10
  & $6.19\times10^{-3}$ (finite) \\
\bottomrule
\multicolumn{4}{l}{\footnotesize $^a$ These values are directly reported or obtained from the experimental data fit at 10dB of channel attenuation.} \\
\label{tab:comparison}
\vspace{-0.5cm}
\end{tabular}
\caption{Comparison with representative recent QKD demonstrations. See main text for details.}
\label{tab:comparison}
\end{table*}

Following the finite-key two-decoy-state protocol of Zhang \emph{et al.}~\cite{Improve2016Ma},  we evaluate our finite-key rate benchmark at a channel attenuation of 10dB, a typical reference loss level used to compare QKD systems.  In our case, the finite-key session uses two-decoy settings optimized for Bob-side's loss of 5.7\,dB. Table~\ref{tab:finitekey} summarizes the measured detection counts, error
counts and the resulting finite-key bounds. Over $N=6.13\times10^9$ emitted pulses, Bob registered $M_z=8.18\times10^7$ detections in the $\mathcal{Z}$ basis. The detection error rates are $e_{\mathrm{det}}^z=1.98\%$ and $e_{\mathrm{det}}^x=2.85\%$. In Fig.~\ref{fig:qberfinite}, we show the QBER distribution, integrated over several blocks of 100 ms, for all the $M_z=8.18\times10^7$ detections registered. As one can see, the QKD scheme works with the QBER distribution stably over the entire session, showing that phase-drift have been properly controlled by the stabilization algorithm. The finite-key rate obtained is 
\begin{equation}
R_{\mathrm{finite}}(10\,\mathrm{dB}) = 6.19\times10^{-3}\ \text{bits/pulse},
\end{equation} which will be compared with modern experiments in the next subsection.

\subsection{Comparison with Previous Works}
\label{subsec:comparison}

Table~\ref{tab:comparison} places the present work in the context of
representative modern QKD demonstrations at a channel attenuation of
10\,dB, a standard reference loss for cross-experiment comparisons.
A central motivation for HD-QKD is the promise of higher noise tolerance
and more information per detected photon relative to qubit-based schemes.
Whether this theoretical advantage translates into competitive secret-key
rates under field conditions has remained an open question, since practical
HD implementations have historically incurred efficiency penalties that
eroded the dimensional gain. The results reported here allow us to examine
this question concretely.

Among HD-QKD demonstrations, the two most directly comparable results are
Islam \emph{et al.}~\cite{Islam2017}, who report
$3.1\times10^{-3}$\,bits/pulse using a $d=4$ time-bin protocol over
spooled SMF with SNSPDs, and Zahidy \emph{et al.}~\cite{Zahidy2024},
who deployed a $d=4$ hybrid core+time-bin system over a 52\,km installed
MCF in L'Aquila. Our result of $6.19\times10^{-3}$\,bits/pulse surpasses Islam \emph{et al.}\ by nearly
a factor of two and Zahidy \emph{et al.}\ by more than a factor of four.
These gains are consistent with the elimination of the encoding penalties
inherent to time-bin based approaches. 

Viewed against the broader QKD landscape, the significance of this result
becomes even clearer. Li \emph{et al.}~\cite{Li2023} recently reported
$8.9\times10^{-3}$\,bits/pulse using a $d=2$ polarization protocol
equipped with a high-speed integrated transmitter and multipixel
SNSPDs. A highly engineered qubit system that currently represents the
state of the art in per-pulse key-generation rate. Our $d=4$ QKD result, reaches $6.19\times10^{-3}$\,bits/pulse (within a factor
of 1.5) of that qubit benchmark. This proximity is notable: it suggests
that, with the encoding efficiency recovered by the pure core-mode
approach, the intrinsic advantages of HD encoding are no longer being
fully absorbed by system overhead. While qubit QKD retains practical
advantages in simplicity and absolute clock-rate scalability, the present
result indicates that HD-QKD is reaching a regime of practical
competitiveness, where its higher information capacity per photon can be
meaningfully harvested in real-world deployments.

\section{Conclusions}
\label{sec:conclusions}

High-dimensional quantum key distribution has received growing attention
as a technique to boost the secret-key rate per protocol round, with
path-encoded implementations over multicore fibers emerging as a
particularly promising route~\cite{Pirandola2020,Xavier2020}. Here we
have demonstrated, for the first time, a $d=4$ HD-QKD protocol based on
pure core-mode encoding over a deployed MCF campus network, operating
under continuous and uncontrolled environmental perturbations. The system
was characterized across three distinct network configurations, ranging
from a back-to-back reference baseline to an installed cross-campus link
spanning 1.3\,km, and benchmarked using both commercially accessible
InGaAs avalanche photodiode detectors and high-efficiency SNSPDs. A key
advance of this implementation is that information is encoded across all
four core modes of the MCF, rather than in a reduced core subspace
supplemented by an auxiliary degree of freedom. This direct encoding
eliminates the generation-rate penalty associated with time-bin schemes
and the photon loss inherent to Franson-like measurement
interferometers~\cite{Alarcon2021}. The use of low-loss multiport beam
splitters fabricated directly in MCF~\cite{Carine2020} for both state
preparation and measurement is central to this result, enabling
high-fidelity qudit operations within the deployed network.

The experiment was performed on the installed MCF network of the
Universidad de Concepci\'on under real-world conditions, including
temperature variations, mechanical perturbations, and
vehicle-induced vibrations along the cross-campus routes. Despite these
perturbations, the common-cladding geometry of the MCF strongly suppresses
differential phase drift between cores~\cite{DaLio2020}, enabling
operational integration times of 100ms, and coherent qudit propagation with mean state fidelity $\bar{F}=0.96\pm0.01$
over the longest deployed link. Positive asymptotic secret-key rates were
obtained over all tested links and with both detector technologies,
confirming viability across a range of operational scenarios. 

To establish a composable finite-key benchmark and enable direct comparison with the
broader QKD literature, a dedicated session was conducted at a reference
channel attenuation of 10\,dB using SNSPDs, yielding a finite-key rate of
$R=6.19\times10^{-3}$\,bits/pulse and a total of $3.80\times10^{7}$
secret key bits from $6.13\times10^{9}$ emitted pulses. This is the
highest per-pulse finite-key rate reported to date for HD-QKD, surpassing
the previous best time-bin result~\cite{Islam2017} by nearly a factor of
two and the best hybrid core+time-bin field demonstration~\cite{Zahidy2024}
by more than a factor of four. Notably, this $d=4$ result approaches
within a factor of 1.5 the current state of the art in qubit
QKD~\cite{Li2023}, suggesting that HD-QKD is entering a regime of genuine
practical competitiveness.

The same architecture is directly scalable: MCFs with larger numbers of
cores and compatible multiport beam splitters~\cite{Carine2020} are
available, allowing higher-dimensional alphabets that further increase
the information encoded per detected photon and improve the noise
tolerance of the protocol. Beyond QKD, the
ability to prepare, transmit, and measure coherent spatial qudits in
deployed multicore fibers is relevant for future metropolitan quantum
networks based on space-division multiplexing~\cite{Xavier2020} and for
quantum communication exploiting high-dimensional spatial
entanglement~\cite{Gomez2021}. Our results thus establish core-mode
encoding as a hardware-efficient and field-viable framework for
high-dimensional quantum-secure communications.

\begin{figure*}[t]
    \centering
    \includegraphics[width=0.9\linewidth]{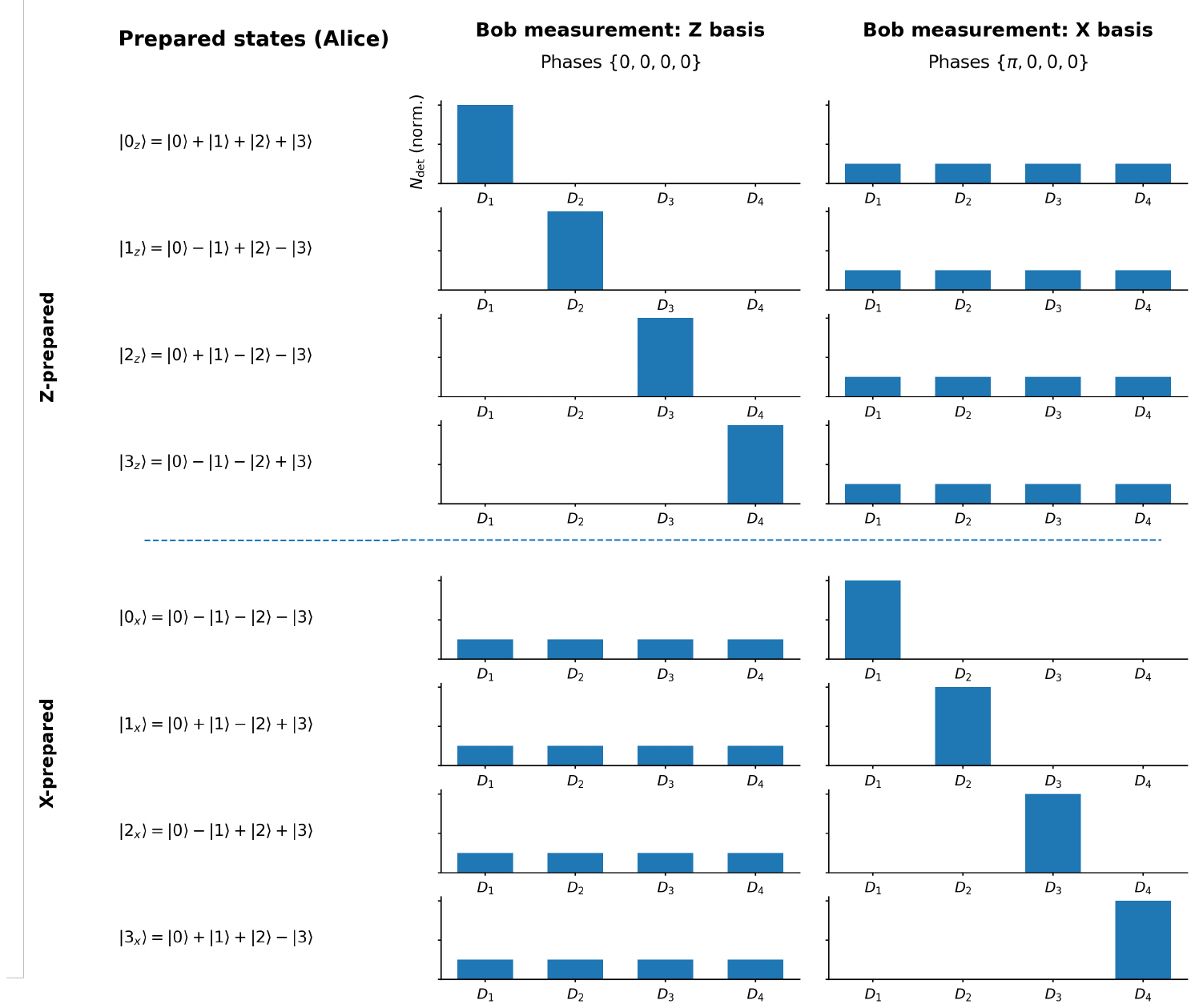}
    \caption{Expected detection distribution for each prepared qudit state at Bob's
    four single-photon detectors ($D_1$--$D_4$), for both the $\mathcal{Z}$
    basis (center column, phase pattern $\{0,0,0,0\}$) and the $\mathcal{X}$
    basis (right column, phase pattern $\{\pi,0,0,0\}$).
    Each row corresponds to one of the eight prepared states.
    A tall bar in detector $D_k$ indicates high probability of detection at
    port $k$; ideally one port dominates for each state, realizing the BB84
    discrimination condition.}
    \label{fig:stateprep}
\end{figure*}

\begin{acknowledgments}
This research was funded by ANID Anillo Project ATE250003, Fondo Nacional de Desarrollo Cient\'{\i}fico y
Tecnol\'ogico (FONDECYT) Grant No.\ 1260111, 1240746, 1240843, 1231940, ANID Millennium Science Initiative Program ICN17\textendash012, ANID AC3E CIA 250006, and Wallenberg Center for Quantum Technologies (WACQT). DM gratefully acknowledges the support from the Austrian Federal Ministry of Labour and Economy, the National Foundation for Research, Technology and the Christian Doppler Research Association.
\end{acknowledgments}

\begin{table*}[h]
\centering
\small
\begin{tabular}{lcccc}
\toprule
Quantity & BtB, $2.16\,\mathrm{dB}$ & P6, $9.36\,\mathrm{dB}$ & PtE, $10.70\,\mathrm{dB}$ & Sim., $14.0\,\mathrm{dB}$ \\
\midrule
    $E_\mu$ & $0.1151 \pm 0.0187$ & $0.1195 \pm 0.0179$ & $0.1417 \pm 0.0255$ & $0.1639 \pm 0.0350$ \\
    $Q_\mu$ & $325.0\times 10^{-5} \pm 5.2\times 10^{-5}$ & $57.1\times 10^{-5} \pm 1.9\times 10^{-5}$ & $46.7\times 10^{-5} \pm 1.9\times 10^{-5}$ & $17.8\times 10^{-5} \pm 1.2\times 10^{-5}$ \\
    $\mu$ & $0.2673 \pm 0.0069$ & $0.2409 \pm 0.0099$ & $0.2658 \pm 0.0124$ & $0.2051 \pm 0.0160$ \\
    $E_\nu$ & $0.1062 \pm 0.0173$ & $0.1402 \pm 0.0237$ & $0.1581 \pm 0.0279$ & $0.2687 \pm 0.0513$ \\
    $Q_\nu$ & $122.4\times 10^{-5} \pm 3.3\times 10^{-5}$ & $23.3\times 10^{-5} \pm 1.3\times 10^{-5}$ & $18.4\times 10^{-5} \pm 1.3\times 10^{-5}$ & $68.6\times 10^{-6} \pm 7.0\times 10^{-6}$ \\
    $\nu$ & $0.0997 \pm 0.0034$ & $0.0936 \pm 0.0060$ & $0.0992 \pm 0.0081$ & $0.0672 \pm 0.0098$ \\
    $e_1^{U}$ & $0.1108 \pm 0.0234$ & $0.1127 \pm 0.0584$ & $0.1233 \pm 0.0752$ & $0.1536 \pm 0.2128$ \\
    $Q_1^{L}$ & $24.5\times 10^{-4} \pm 1.9\times 10^{-4}$ & $43.4\times 10^{-5} \pm 9.3\times 10^{-5}$ & $34.4\times 10^{-5} \pm 9.2\times 10^{-5}$ & $13.4\times 10^{-5} \pm 7.8\times 10^{-5}$ \\
    $R_\infty$ & $8.77\times 10^{-4} \pm 5.00\times 10^{-4}$ & $1.64\times 10^{-4} \pm 1.40\times 10^{-4}$ & $6.04\times 10^{-5} \pm 1.35\times 10^{-4}$ & $8.62\times 10^{-6} \pm 7.63\times 10^{-5}$\\
\bottomrule
\end{tabular}
\caption{Recorded data used for the asymptotic calculations with the InGaAs SPDs. Pdark = $3.81 \times 10^{-6} \pm 2.27 \times 10^{-6}$.}
\label{tab:app_spad_mean_inputs}
\end{table*}

\begin{table*}[h]
\centering
\small
\begin{tabular}{lcccc}
\toprule
Quantity & BtB, $2.16\,\mathrm{dB}$ & P6, $9.36\,\mathrm{dB}$ & PtE, $10.70\,\mathrm{dB}$ & Sim., $20.0\,\mathrm{dB}$ \\
\midrule
    $E_\mu$ & $0.0247 \pm 0.0269$ & $0.0510 \pm 0.0015$ & $0.0392 \pm 0.0014$ & $0.0447 \pm 0.0059$ \\
    $Q_\mu$ & $603.0\times 10^{-4} \pm 7.5\times 10^{-4}$ & $1897.1\times 10^{-5} \pm 2.1\times 10^{-5}$ & $1227.2\times 10^{-5} \pm 1.5\times 10^{-5}$ & $1816.1\times 10^{-6} \pm 7.9\times 10^{-6}$ \\
    $\mu$ & $0.8685 \pm 0.0106$ & $0.7296 \pm 0.0073$ & $0.6657 \pm 0.0067$ & $0.7807 \pm 0.0089$ \\
    $E_\nu$ & $0.0603 \pm 0.0129$ & $0.0743 \pm 0.0059$ & $0.1320 \pm 0.0104$ & $0.3567 \pm 0.0358$ \\
    $Q_\nu$ & $189.5\times 10^{-5} \pm 9.7\times 10^{-5}$ & $1187.0\times 10^{-6} \pm 5.3\times 10^{-6}$ & $709.8\times 10^{-5} \pm 5.8\times 10^{-6}$ & $137.8\times 10^{-6} \pm 4.8\times 10^{-6}$ \\
    $\nu$ & $0.0134 \pm 0.0007$ & $0.0433 \pm 0.0005$ & $0.0341 \pm 0.0005$ & $0.0371 \pm 0.0032$ \\
    $e_1^{U}$ & $0.0411 \pm 0.0146$ & $0.0463 \pm 0.0075$ & $0.0870 \pm 0.0135$ & $0.1248 \pm 0.0813$ \\
    $Q_1^{L}$ & $0.0497 \pm 0.0039$ & $90.4\times 10^{-4} \pm 1.4\times 10^{-4}$ & $65.0\times 10^{-4} \pm 1.4\times 10^{-4}$ & $8.1\times 10^{-4} \pm 1.1\times 10^{-4}$ \\
    $R_\infty$ & $5.96\times 10^{-2} \pm 2.33\times 10^{-2}$ & $7.63\times 10^{-3} \pm 5.01\times 10^{-4}$ & $5.36\times 10^{-3} \pm 4.98\times 10^{-4}$ & $4.34\times 10^{-4} \pm 3.26\times 10^{-4}$\\
\bottomrule
\end{tabular}
\caption{Recorded data used for the asymptotic calculations with the SNSPDs. Pdark = $13.50 \times 10^{-6} \pm 1.30 \times 10^{-6}$.}
\label{tab:app_snspd_inputs}
\end{table*}

\appendix

\section{Finite-Key Parameter Estimation Procedure}
\label{app:chernoff}

Here we present the complete procedure used to estimate
the finite-key bounds $M_1^{zL}$, $M_0^{zL}$, and $e_1^U$ appearing in
Eq.~\eqref{eq:finiteK} of the main text.
The analysis follows the method of Zhang \emph{et al.}~\cite{Improve2016Ma}
for the two-decoy finite-key protocol. We adapt
the method to the high-dimensional case and extend it to obtain a
finite-key correction for the bound $M_0^{zL}$.

\subsection*{Step 1: Inverse Chernoff bounds and $Y_1^{*L}$}

For any experimentally observed count $\chi$, the quantities
$\mathbb{E}^L[\chi]$ and $\mathbb{E}^U[\chi]$ denote, respectively,
lower and upper bounds on the expected value $\mathbb{E}[\chi]$ obtained
using inverse Chernoff bounds. These bounds are given by
\begin{align}
\mathbb{E}^L[\chi] &= \frac{\chi}{1+\delta^L}, &
\mathbb{E}^U[\chi] &= \frac{\chi}{1-\delta^U}, \label{eq: A1}
\end{align}

where $\delta^L$ and $\delta^U$ are obtained numerically from
\begin{align}
\left[ \frac{e^{\delta^L}}{(1+\delta^L)^{1+\delta^L}}\right]^{\chi/(1+\delta^L)}
  &= \tfrac{1}{2}\epsilon, \\
\left[ \frac{e^{-\delta^U}}{(1-\delta^U)^{1-\delta^U}}\right]^{\chi/(1-\delta^U)}
  &= \tfrac{1}{2}\epsilon,
\end{align}
for a chosen failure probability $\epsilon = 10^{-10}$.

Using $\mathbb{E}[M_a]/N_a = \mathbb{E}[Q_a]$, where $Q_a$ denotes the
gain associated with WCPs of type $a$, the finite-key analogue of the single-photon
yield lower bound is

\begin{align}
Y_1^{*L} = &\frac{\mu}{\mu\nu - \nu^2}\Bigl[
  \frac{\mathbb{E}^L[M_\nu]}{N_\nu} e^\nu \nonumber
  \\
  &- \frac{\nu^2}{\mu^2}\frac{\mathbb{E}^U[M_\mu]}{N_\mu}  e^\mu
  - \frac{\mu^2-\nu^2}{\mu^2} \frac{\mathbb{E}^U[M_0]}{N_0}  \Bigr].
\label{eq:Y1starL}
\end{align}

The corresponding lower bound on the expected number of single-photon
detections in the $\mathcal{Z}$ basis across all WCPs is then
\begin{align}
M_1^{*zL} = Y_1^{*L}\, Np_z^2\, \bigl( p_\mu e^{-\mu}\mu + p_\nu e^{-\nu}\nu \bigr).
\label{eq:M1starL}
\end{align}

\subsection*{Step 2: Lower bound $M_1^{zL}$}

The quantity $M_1^{*zL}$ provides a bound on the expected number of
single-photon detections across all WCPs.
To obtain the bound restricted to the signal WCP,
the symmetric Chernoff bound is applied with
$\bar\chi = \mathbb{E}[M_1^{zL}] = p_{1s}\,M_1^{*zL}$,
where $p_{1s}$ denotes the probability that a single-photon detection in
the $\mathcal{Z}$ basis is from the signal WCP.
This yields
\begin{align}
M_1^{zL} = (1-\delta)\,\bar\chi,
\label{eq:Msz1L}
\end{align}
with
\begin{equation}
\delta = \frac{-\ln(\epsilon/2)
  + \sqrt{\ln(\epsilon/2)^2 - 8\ln(\epsilon/2)\,\bar\chi}}{\bar\chi}.
\label{eq:simcher}
\end{equation}

\subsection*{Step 3: Lower bound $M_0^{zL}$}

An analogous inverse-Chernoff analysis is applied to the vacuum detections.
This gives
\begin{equation}
Y_0^{*L} = \frac{\mathbb{E}^L[M_0]}{N_0};
\end{equation}
\begin{align}
M_0^{*zL} = Y_0^{*L}\, Np_z^2\,
  \bigl( p_\mu e^{-\mu} + p_\nu e^{-\nu} + p_v \bigr),
\label{eq:M0starL}
\end{align}
from which the bound on vacuum detections associated with the signal
intensity is obtained as
$M_0^{zL} = (1-\delta)\,\bar\chi$ with
$\bar\chi = p_{0s}\,M_0^{*zL}$,
where $p_{0s}$ denotes the probability that a vacuum detection in the
$\mathcal{Z}$ basis came from the signal WCP. \\

\subsection*{Step 4: Upper bound $e_1^U$ from $\mathcal{X}$ basis}

Inverse Chernoff bounds applied to the error counts yield
\begin{align}
e_1^{xU} =
\frac{\frac{\mathbb{E}^U[E_{\nu x} M_{\nu x}]}{N_{\nu x}}\,e^\nu
      - \frac{\mathbb{E}^L[E_{0x}M_{0x}]}{N_{0x}}}{Y_1^{*L}\,\nu}.
\label{eq:e1ux}
\end{align}

To relate this quantity to the phase-error rate in the $\mathcal{Z}$ basis,
a random-sampling argument is used.
The correction $\theta$ satisfying $e_1^U = e_1^{xU} + \theta$ is obtained
numerically from
\begin{align}
\epsilon =
\frac{\sqrt{M_1^{*xL}+M_1^{zL}}}
     {\sqrt{e_1^{xU}(1-e_1^{xU})\,M_1^{*xL}\,M_1^{zL}}}\,
2^{-(M_1^{*xL}+M_1^{zL})\,\varepsilon(\theta)},
\label{eq:randomsampling}
\end{align}
where
$\varepsilon(\theta) = H_2(e_1^{xU}+\theta - q_x\theta)
  - q_x H_2(e_1^{xU}) - (1-q_x) H_2(e_1^{xU}+\theta)$
and $q_x = M_1^{*xL}/(M_1^{*xL}+M_1^{zL})$.

Substituting the resulting bounds $M_1^{zL}$, $M_0^{zL}$, and $e_1^U$
into Eq.~\eqref{eq:finiteK} yields the composable secret key length $K$,
guaranteeing $\epsilon_{\sec}$ security and $\epsilon_{\mathrm{cor}}$
correctness against general attacks.

In the case that the parameters $\epsilon_{\mathrm{cor}}, \epsilon_{\mathrm{sec}} > 0$,
the protocol is said to be
$(\epsilon_{\mathrm{cor}}+\epsilon_{\mathrm{sec}})$-secure \cite{Lim14}.
Correctness is satisfied if
\begin{equation}
\Pr[K_A \neq K_B] \leq \epsilon_{\mathrm{cor}},
\end{equation} where $K_A$ and $K_B$ denote the final secret keys held by Alice and Bob,
respectively. Thus, the probability that Alice and Bob obtain different final keys is bounded by $\epsilon_{\mathrm{cor}}$. Secrecy is satisfied if
\begin{equation}
(1-p_{\mathrm{abort}})
\left\|
\rho_{AE} - U_A \otimes \rho_E
\right\|_{1/2}
\leq
\epsilon_{\mathrm{sec}},
\end{equation} where $\rho_{AE}$ is the joint state
of Alice's final key and Eve's system. $U_A$ is the uniform mixture over all
possible values of Alice's final key, and $p_{\mathrm{abort}}$ is the
probability that the protocol aborts. This condition guarantees that, except with probability
$\epsilon_{\mathrm{sec}}$, the final key is indistinguishable from an ideal
uniform secret key independent of Eve's information. 

In practical QKD implementations, the correctness parameter is commonly
chosen as $\epsilon_{\mathrm{cor}} = 10^{-15}$, while the secrecy parameter
is typically taken to scale with the final secret-key length. Following the
standard convention adopted in the literature, the secrecy contribution is
therefore written as $\epsilon_{\mathrm{sec}} = 10^{-15} \times K$, prior to the
additional finite-key corrections introduced by the statistical estimation
procedure.


\section{{$4$-dimensional BB84 QKD states and basis}}
\label{app:d4BB84}

In Fig.~\ref{fig:stateprep} we give the explicit form of the states prepared by Alice and measured by Bob.


\section{{Recorded data used in the asymptotic key rate calculations}}
\label{app:data}

In Table~\ref{tab:app_spad_mean_inputs} and Table~\ref{tab:app_snspd_inputs}, we list the recorded data used to reproduce all the asymptotic key-rate calculations reported in the main text with the InGaAs SPDS and SNSPDs, respectively.

\bibliographystyle{apsrev4-2}
\bibliography{main}

\end{document}